# Signature of non-exponential nuclear decay

A.Ray[1], A. K. Sikdar[1], A. De[2]

[1]*Variable Energy Cyclotron Centre,1/AF Bidhannagar, Kolkata-700064, India.*

[2]*Raniganj Girls' College, Raniganj, West Bengal, India.*

Precision tests of decay law of radioactive nuclei have not so far found any deviation from the exponential decay law at early time, as predicted by quantum mechanics. In this paper, we show that the quantum decoherence time (i.e. the timescale of non-exponential decay) of the quasifission/fission process should be $\sim 10^{-18}$ sec considering the atom of the fissioning nucleus as a quantum detector. Hence, the observed decay timescale of the quasifission/fission process of even highly excited ($E_X > 50$ MeV) transuranium and uranium-like complexes should be rather long ($\sim 10^{-18}$ sec) in spite of their very fast exponential decay timescale ($\sim 10^{-21}$ sec - $10^{-20}$ sec) as measured by the nuclear techniques. Recent controversy regarding the observation of very long ($\sim 10^{-18}$ sec) and very short ($\sim 10^{-21}$ sec – $10^{-20}$ sec) quasifission/fission timescales for similar systems at similar excitation energies as obtained by direct techniques (crystal blocking, X-ray-fission fragment) and nuclear techniques could be interpreted as evidence for non-exponential decays in nuclear systems.

The exponential decay law is the hallmark of all radioactive decays studied so far. However, according to quantum mechanics [1-3], there should be deviations from the exponential decay law both at very early and later times. The unitary evolution of an unstable state cannot lead to exponential decay and the onset of exponential decay is related to the loss of quantum coherence between the initial parent state and the decayed state. So, there must exist a region between t=0 instant and the known exponential decay domain where the decay should be non-exponential. Short time deviation from exponential decay was observed earlier in the case of quantum tunneling of ultra-cold sodium atoms from a periodic optical potential created by a standing wave of light [4]. However no such deviation from the exponential decay law has been observed so far for radioactive decays. High precision tests of the decay law of radioactive nuclei at early time were carried out by Norman et al. [5,6] and they did not find any deviation from the exponential decay law down to $10^{-4} t_{1/2}$ time for $^{60}$Co and $10^{-10} t_{1/2}$ time for $^{40}$K nuclei. Although, Norman et al. monitored decay curve at an early time that is a very small fraction of the half-life of the radioactive nucleus, they actually started monitoring the decay curve a few hours after the formation of the radioactive nuclei and this timescale of a few hours is too long [7] to observe non-exponential decay of the radioactive nuclei. In nuclear decays, the timescale of exponential decay is generally much longer than that of the non-exponential decay that takes place at an early time inaccessible to direct monitoring of the decay curve. However, the timescale of exponential decay for nuclear systems does not always have to be much longer than the corresponding non-exponential timescale. The reverse might also be true in some situations. We shall show in this paper that in the case of very fast quasifission/fission processes of highly excited ($E_X > 50$

MeV) transuranium and uranium-like nuclei, the non-exponential timescale could be much longer than the exponential timescale resulting in unusually long observed timescale of the decay process and providing evidence for the existence of non-exponential decay at early time.

Recently, R. du Rietz et al. [8] measured mass-angle distributions of fragments from the reactions $^{64}$Ni+W at E($^{64}$Ni)$_{Lab}$= 310-341 MeV, $^{48}$Ti+W at E($^{48}$Ti)$_{Lab}$=220-260 MeV and $^{34}$S+W at E($^{34}$S)$_{Lab}$=149-189 MeV. They performed simulations of the mass-angle distributions of fragments using a classical rotational nuclear model, parametrizing nuclear sticking time distribution with a half Gaussian followed by an exponential decay function. The sticking time was converted to the observed scattering angle using calculated angular momentum and moment of inertia of the system. Comparing the simulation with the corresponding experimental mass-angle distribution of fragments, they obtained that the timescales for $^{64}$Ni+W and $^{48}$Ti+W quasifission reactions were $5\times10^{-21}$ sec and $10\times10^{-21}$ sec respectively, whereas the timescale of $^{34}$S+W fission reaction was found to be greater than $10^{-20}$ sec. R. du Rietz et al. [8] actually extracted the exponential decay time of the quasifission process. In order to show it explicitly, we have used their relevant 2D color scatter plots given in (ref. 8) and made plots (as shown in Fig. 1) of $\left(\frac{d\sigma}{d\theta}\right)_{c.m.}$ versus $\theta_{c.m}$ for $M_F$=0.3 and 0.4, where $M_F$ is the ratio of the fragment mass to the sum of projectile and target mass for the reaction $^{64}$Ni+W at E($^{64}$Ni)$_{Lab}$=341 MeV. The observed exponentially dropping angular distribution $\left(\frac{d\sigma}{d\theta}\right)_{c.m.} \propto e^{-\frac{\theta_{c.m.}}{\gamma}}$, (where $\theta_{c.m.}$ is the scattering angle in the center of mass frame) as shown in Fig. 1, implies [9] an exponential decay law $\sigma(\theta) \propto \frac{dN(t)}{dt} \propto e^{-\lambda t}$, where N(t) is the number of dinuclear complexes at the instant t and the mean lifetime of the exponential decay is $\tau = \frac{1}{\lambda} = \left(\frac{I}{L}\right)\gamma$, where I, L and γ are the moment of inertia, orbital angular momentum of the dinuclear system and damping constant respectively. Using the values of γ from the observed slopes of the exponential angular distributions (Fig. 1) and the calculated values [10] of I and L, we find that the values of dinuclear lifetime (τ) increases from $10^{-21}$ sec to $4.4 \times 10^{-21}$ sec as $M_F$ increases from 0.25 to 0.45 for $^{64}$Ni+W reaction at E($^{64}$Ni)$_{Lab}$=341 MeV, in good agreement with the results of R. du Rietz et al [8]. Similar results and agreements were obtained for other reactions studied by R. du Rietz et al. [8].

Hinde et al. [11] and Ramachandran et al. [12] produced uranium-like and transuranium nuclei at high excitation energies and measured their fission lifetime of the order of $10^{-20}$ sec by prescission neutron multiplicity technique. Hinde et al. [11] produced Z ≈120, A ≈302 nuclei by $^{64}$Ni+$^{238}$U reaction at $E_{c.m.}$ = 329 MeV and measured fission lifetimes of the order of $10^{-20}$ sec by prescission neutron multiplicity technique and Toke et al. [9] measured sticking time of $^{238}$U+$^{64}$Ni nuclei at similar excitation energy as $7.5 \times 10^{-21}$ sec. On the other hand, non-nuclear crystal blocking [13-15] and X-ray coincidence techniques [16-18] measured lifetime (that should include both the non-exponential and exponential decay timescales) of the quasifission/fission processes by less model-dependent direct approaches. Hence, if the timescale of non-exponential decay could be ignored, the lifetimes of quasifission/fission processes obtained from nuclear techniques should be consistent with those obtained from the crystal blocking and X-ray coincidence techniques for similar systems at similar energies.

However, it was found that both the crystal blocking [13-15] and X-ray techniques [16-18] obtained lifetimes of the order of $10^{-18}$ sec for both the quasifission and fission processes even at high excitation energies (>50 MeV) for the transuranium and uranium-like nuclei. Andersen et al. [13] measured the timescales of quasifission reactions $^{74}$Ge+W at E($^{74}$Ge)$_{Lab}$=390 MeV, $^{58}$Ni+W at E($^{58}$Ni)$_{Lab}$=330-375 MeV, $^{48}$Ti+W at E($^{48}$Ti)$_{Lab}$=240-255 MeV and that of fission reaction $^{32}$S+W at E($^{32}$S)$_{Lab}$=180 MeV by crystal blocking technique and obtained about one attosecond ($10^{-18}$ sec) for the timescales of all those quasifission and fission processes. They found [13] that all the observed quasifission and fission fragments came from slow processes of lifetime about an attosecond. Molitoris et al. [16] found from their X-ray-fission fragment coincidence measurement that the fission lifetime of uranium-like nuclei (even at excitation energy $E_X$=105 MeV) was $\geq 4\times10^{-18}$ sec and most of the fission fragments (≥52%) came from such slow fission process. Fregeau et al. [17] produced Z=120 nuclei at high excitation ($E_X > 50$ MeV) energies by $^{238}$U+$^{64}$Ni reaction at $E_{c.m.} = 333$ MeV and found that most of the fission fragments (> 53%) that they detected (70≤ Z ≤80) came from a slow fission process of lifetime $> 2.5\times10^{-18}$ sec. So, these experiments indicate that all or most of the quasifission/fission fragments come from slow processes of lifetime $> 10^{-18}$ sec. Hence, there exists a large discrepancy between the absolute values of the quasifission and fission timescales obtained from nuclear and direct techniques.

It might be possible to explain the observed attosecond (~$10^{-18}$ sec) quasifission/fission lifetime of highly excited uranium and transuranium nuclei by increasing the value of reduced friction parameter [19,20] due to nuclear viscosity and for Z=120 nuclei [15,17] by using the concept of super-heavy nuclei having large fission barrier. However, the primary discrepancy is that the quasifission/fission lifetimes measured for similar systems at similar excitation energies by nuclear and direct techniques give results that are orders of magnitude different as discussed before. Using very large values of friction parameter, Jacquet and Morjean [20] showed calculations of very broad fission time distributions extending to $10^{-15}$ sec and pointed out that the nuclear techniques were sensitive to short timescales ($10^{-21}$ sec – $10^{-19}$ sec), thus attempting to explain the anomaly. If we assume a very broad quasifission/fission time distribution implying that most of the quasifission/fission fragments should be coming from a very slow process, mass-angle distribution technique should have seen, on the average, very long sticking time ($>> 10^{-20}$ sec) for all the reactions studied, contrary to the observations [8, 9]. Very broad fission time distributions [19,20] are not consistent with prescission neutron multiplicity data from the highly excited uranium and transuranium nuclei [11,12]. Hinde et al. [11] produced highly excited ($E_X \approx 240$ MeV) uranium-like and transuranium nuclei and after the emission of all the observed prescission neutrons, they had average excitation energy of around 190 MeV as they fissioned. Since neutron emission from uranium and transuranium nuclei at $E_X \approx 200$ MeV should be a very fast process, Hinde et al. [11] deduced fission lifetime $\sim 10^{-20}$ sec for those highly excited uranium and transuranium nuclei. If we assume a very broad fission time distribution [19, 20] extending to $10^{-15}$ sec time scale, neutron emission from the highly excited uranium and transuranium nuclei should have continued to much lower excitation energy, thus emitting a very large number of prescission neutrons and contradicting observations [11]. So, the timescale (~$10^{-20}$ sec) obtained for highly excited uranium and transuranium nuclei using nuclear techniques are not consistent with very broad quasifission/fission time distributions [19, 20]. Jacquet and Morjean showed [20] a calculation of fission time distribution for $^{58}$Ni+$^{208}$Pb reaction at 8.86 MeV/A where the time distributions for quasifission and fission processes peaked around $10^{-20}$ sec and $6 \times 10^{-20}$ sec respectively and essentially no events were seen at ~$10^{-18}$ sec timescale. Such a time distribution would imply that the experiments using direct techniques would hardly see any events, contradicting observations [13-18]. Hence, we cannot explain the situation by the sensitivity of the techniques to different timescales or by fission/quasifission time distribution plots.

The observed lifetime of quasifission/fission process by any direct method (crystal blocking or X-ray technique) could be significantly longer than the lifetime of the exponential decay, if the non-exponential decay is much slower than the exponential decay. Let us make an order of magnitude estimate of the quantum decoherence time of the nuclear quasifission/fission process. The quantum decoherence process of the nuclear quasifission/fission process might be thought as a two step process [21, 22]. In the first step, the entire ion whose nucleus is undergoing fission acts as a quantum detector observing the nuclear fission/quasifission process. As a result, after a certain time, the nuclear system couples with all the atomic orbitals of the ion and produces a fully correlated pure nuclear-ion state. So, the diagonal elements of the corresponding density matrix will be real numbers and the off-diagonal elements will contain complex numbers expressing purely quantum correlations. In the second step, by considering the interaction of the environment with the pure nuclear-ion state, one gets a reduced density matrix by tracing over the environment [21, 22] and this reduced density matrix contains only classical correlations, thus indicating complete loss of quantum coherence. After the loss of quantum coherence, classical descriptions and exponential decay law for fission/quasifission process are applicable. The time evolution of the unstable decaying dinuclear state $(|\psi_{dinuclear}>)$ should produce the following superposition of states:

$$e^{-\frac{iHt}{\hbar}}|\psi_{dinuclear}> = \sum_i \alpha_i(t)|\psi_{fissioned}>_i + \beta(t)|\psi_{dinuclear}> \quad (1)$$

where $|\psi_{fissioned}>_i$ indicate different macroscopically distinguishable fission fragment pairs. These states should be orthogonal to one another i.e. $\langle(\psi_{fissioned})_i|(\psi_{fissioned})_j\rangle = 0$ for $i \neq j$ and $\langle(\psi_{fissioned})_i|\psi_{dinuclear}\rangle = 0$. It can be shown [1,2] that

$$\beta(t+t') = \beta(t)\beta(t') + \sum_i \alpha_i(t)\langle\psi_{dinuclear}|\exp(-iHt')|\psi_{fissioned}\rangle_i \quad (2)$$

and $\beta(t) \propto \exp(-\lambda t)$ with $Re(\lambda) > 0$, only when the reformation amplitude of the unstable state $\psi_{dinuclear}$ from the fission fragments (as given in eq. (2)) becomes zero. However, it can never be zero from the solution of the time dependent Schrodinger equation. Hence, the condition for the start of the exponential decay and the classical description of the process is the loss of the quantum coherence of the nuclear states as a result of the interaction of the system with the environment.

Vacancies in the atomic orbitals of the target atom are created when a projectile ion collides with a target atom to produce a compound atom with a fused compound nucleus. The excited combined ion might be represented by the following coherent superposition of all the atomic orbitals: $|A^*> = \sum_{i=1}^{i=n} C_i A_i^*$, where $A_i^*$ indicates atomic wave function with one vacancy in $i^{th}$ orbital and n is total number of orbitals. The magnitude of $C_i$ increases rapidly for outer electronic orbitals [23]. Let us consider for simplicity a superposition of two most likely states as a result of the time evolution of the dinuclear state

$$e^{-\frac{iHt}{\hbar}}|\psi_{dinuclear}> \approx \alpha_p(t)|\psi_{fissioned}>_p + \beta(t)|\psi_{dinuclear}>, \text{ where } |\psi_{fissioned}>_p$$

indicates the most probable fission fragment pair and $\left\langle \left(\psi_{fission}\right)_p \middle| \psi_{dinuclear} \right\rangle = 0$.

$\alpha_p(t)$ and $\beta(t)$ are complex coefficients. Let us now consider the coupling of the nuclear state with the ionic state $|A^*>$, where $|A^*> = \sum_{i=1}^{i=n} C_i A_i^*$. The interactions of the dinuclear state and fissioned state with the atomic orbital state $|A_i^*>$ containing a vacancy in $i^{th}$ orbital might be represented as follows:

$$|\psi_{dinuclear}>|A_i^*> \rightarrow \gamma_i(t)|\psi_{dinuclear}>|A_i, X_i> \tag{3}$$

where $\gamma_i(t) = 1 - e^{-\lambda_i t}$ and $\lambda_i$ is the decay rate of $i^{th}$ vacancy as it is filled up from the higher orbitals. Here, $|A_i, X_i>$ denotes an ion with no vacancy in $i^{th}$ orbital, but a vacancy in the higher orbital due to the electronic transition and $X_i$ denotes the corresponding X-ray photon. Regarding the interaction of $|\psi_{fission}>_p$ with $|A_i^*>$, at a very early time, the distant atomic orbital will not experience any significant effect as the fission fragments start separating out. After a certain time $t_i$, as the separation between the fission fragments becomes comparable to the diameter of the $i^{th}$ atomic orbital, the electronic configuration becomes so much altered that the overlap of the electronic wave function with the original $i^{th}$ orbital becomes negligible. So the $i^{th}$ orbital might be considered destroyed and characteristic photon emission due to the transition to the $i^{th}$ orbital cannot take place. We indicate the state with destroyed $i^{th}$ orbital and no corresponding photon emission ($X_i$) due to the transition to the $i^{th}$ orbital as $|\bar{A}_i, \bar{X}_i>$.

This state has been considered orthogonal to $|A_i, X_i>$ i.e. $\langle A_i, X_i | \bar{A}_i, \bar{X}_i \rangle = 0$. So for $t \ll t_i$, the interaction of both $|\psi_{fission}>_p$ and $|\psi_{dinuclear}>$ on $|A_i^*>$ should be very similar. Hence for $t \ll t_i$,

$$|\psi_{fission}>_p |A_i^*> \rightarrow \gamma_i(t)|\psi_{fission}>_p |A_i, X_i> \tag{4}$$

However for $t \geq t_i$,

$$|\psi_{fission}>_p |A_i^*> \rightarrow |\psi_{fission}>_p |\bar{A}_i, \bar{X}_i> \tag{5}$$

We have the orthogonality conditions $\langle A_i, X_i | A_j, X_j \rangle = 0$ for i≠j. So for $t \geq t_n$, (where the total number of atomic orbitals =n), the combined dinuclear-ion state evolves [21, 22] into a correlated state

$$|\psi_{correlated}> = \alpha_p(t) \sum_{i=1}^{n} C_i |\psi_{fission}>_p |\bar{A}_i, \bar{X}_i>$$
$$+ \beta \sum_{i=1}^{n} C_i \gamma_i |\psi_{dinuclear}> |A_i, X_i> \tag{6}$$

where the summation is over all the atomic orbitals. The corresponding density matrix of the pure state $|\psi_{correlated}>$ is $\rho^c = |\psi_{correlated}> <\psi_{correlated}|$ whose diagonal elements are real numbers. After tracing over the environment [21, 22], one gets the reduced density matrix

$$\rho^r = |\alpha_p|^2 \sum_{i=1}^{n} |C_i|^2 |\psi_{fission}>_p <\psi_{fission}|_p |\bar{A}_i, \bar{X}_i> <\bar{A}_i, \bar{X}_i|$$
$$+ |\beta|^2 \sum_{i=1}^{n} |C_i|^2 |\gamma_i|^2 |\psi_{dinuclear}> <\psi_{dinuclear}| |A_i, X_i> <A_i, X_i|$$

The reduced density matrix contains only classical correlations implying that the system would be either in the fissioned state or dinuclear state. Since an ion is a much bigger object than a nucleus, decoherence time of the nuclear-ion system due to its interaction with the environment (i.e. tracing over the environment) should be much faster [22] than decoherence time of the fission/quasifission process. So, we might consider the time required to transform from $\rho^c$ to $\rho^r$ as instantaneous and take the time $t_n$ required to form the coupled state with all the orbitals of the ion (as given in eq. (6)) as the decoherence time of the fissioning nucleus. It is very important to couple with all the atomic orbitals, because otherwise the reduced density matrix will contain quantum correlations and decoherence will not be achieved. For example, at $t = t_{i'}$ ($t_{i'} \ll t_n$, n being the total number of atomic orbitals), the nuclear-ion wave function might be approximately written as

$$|\psi_{corre\_partial}> \approx$$

$$\alpha_p(t) \left[ \sum_{i=1}^{i'} C_i |\psi_{fission}>_p |\bar{A}_i, \bar{X}_i> + \sum_{i=i'}^{n} C_i \gamma_i(t) |\psi_{fission}>_p |A_i, X_i> \right]$$

$$+ \beta \sum_{i=1}^{n} C_i \gamma_i |\psi_{dinuclear}> |A_i, X_i>$$

The diagonal elements of the corresponding density matrix $\rho_{corre\_partial} = |\psi_{corre\_partial}><\psi_{corre\_partial}|$ are complex numbers and after tracing over the environment, the reduced density matrix will contain quantum correlations. However, for $t \geq t_n$, the fully correlated coupled state as given by eq. (6) would be produced and after tracing over the environment, the reduced density matrix will contain only classical correlations and quantum decoherence would be achieved.

So, the question is what is the estimate for $t_n$? The problem of two approaching atoms forming a united atom was reviewed by J. Reinhardt and W. Greiner [24]. In our case, the nucleus of the united atom is breaking apart due to fission. Since the speed (~$10^9$ cm/sec) of the fission fragments is much smaller than the orbital speed of the electrons, adiabatic approximation could be applied. As the fission fragments start separating out, the electronic wave function is described in terms of molecular orbitals [25] involving two separating charge centers. However when the distance between the charge centers is very small compared to the diameter of $i^{th}$ atomic orbital of the compound atom, the overlap of the $i^{th}$ orbital with the electronic wave function would be large. As the distance between the charge centers becomes comparable to the diameter of the $i^{th}$ orbital the overlap of the electronic wave function with the original $i^{th}$ orbital becomes negligible and the $i^{th}$ atomic orbital of the compound atom could be considered destroyed. Hence when the distance between the charge centers is comparable to the diameter of the atom/ion, all the atomic orbitals could be considered destroyed. An estimate of the upper limit of the destruction time of all the atomic orbitals might be done by calculating the time taken by the fission fragments to cross the atomic radius i.e. $\frac{10^{-8} cm}{10^9 cm/sec} = 10^{-17}$ sec. So $t_n < 10^{-17}$ sec i.e. $t_n$ should be of the order of $10^{-18}$ sec. The classical exponential decay of the nuclear system should start after the loss of quantum coherence in attosecond time scale and very little fission decay would occur before that. Hence, only direct techniques would be able to measure the total decay time i.e. the sum of quantum decoherence time and the classical exponential fission/quasifission decay time $\approx$ (~$10^{-18}$ sec + classical quasifission/fission decay time). At the lower excitation energy of the compound nucleus, as the classical exponential fission decay time would be much longer than $10^{-18}$ sec, the fission time would be dominated by the classical exponential decay time and so it could be much longer than $10^{-18}$ sec. However at higher excitation energy for uranium and transuranium nuclei, the classical quasifission/fission time could be of the order of $10^{-21}$ sec to $10^{-20}$ sec and the lower limit of the fission decay time would be set by the quantum decoherence time. The quantum decoherence time is expected to be shorter at the higher excitation energy of the compound nucleus for two reasons. 1) At higher excitation energy, more energetic fission fragments moving at higher speeds are emitted, thus reducing the destruction time of the orbitals slightly. 2) If highly excited compound nuclei are produced by smaller impact parameter collisions [14], they would form ions of higher charge states with smaller ionic radii, thus reducing the quantum decoherence time.

Qualitatively speaking, the interaction of the dinuclear/compound state with the atomic states producing photons is like a measurement process on the dinuclear/compound state starting from t=0 instant when the dinuclear state was formed, indicating the presence of the dinucleus or compound nucleus. This measurement process is creating quantum Zeno effect [2] inhibiting the time evolution of the dinuclear state. We consider the timescale of creation of fully correlated nuclear-ion wave function as decoherence time of the quasifission/fission process and this is the timescale of destruction of all the atomic orbitals when no further photon emission due to the transitions of electrons between orbital states is possible.

The time dependent Hartree-Fock calculations (TDHF) [26, 27] apparently show that the quasifissioning fragments separate out in zeptosecond ($\sim 10^{-21}$ sec) timescale. However, in TDHF calculations of heavy ion collisions, the relative motion between the nuclei is treated classically while the internal degrees of freedom of nucleons and their couplings to collective excitations are treated quantum mechanically on a mean field level described by a single Slater determinant. TDHF calculations consider time evolution of an initial state wave function that is a product of ground state wave functions of two nuclei boosted with a relative velocity. After collision, single particle wave functions extend spatially and the final stage wave function remains a superposition of states with different particle number distributions. Classical exponential decay of the dinuclear system should not take place as long as the superposition of dinuclear and fissioned states persists. However, TDHF calculations use classical trajectories and for collisions outside the fusion critical impact parameter, the fragments are assumed to separate out irreversibly, although the final state wave function remains a superposition of states. In other words, quantum decoherence time is not included in TDHF calculations and the separation time of the fragments is essentially based on classical considerations.

In summary, the controversy [8] regarding the quasifission/fission timescale is probably because the nuclear techniques such as mass-angle distributions measure the exponential decay time of the dinuclear system after the loss of the quantum coherence of the superposed nuclear states, whereas crystal blocking/X-ray experiments measure total lifetime that is essentially a sum of the quantum decoherence and classical exponential decay times. TDHF codes [26, 27] give classical contact time of the two nuclei on the basis of trajectory calculations and quantum decoherence time is not included in any such calculations. We conclude from our quantum mechanical analysis that the observed long quasifission/fission timescale ($\sim 10^{-18}$ sec) measured by direct methods such as crystal blocking and X-ray techniques imply non-exponential decay of the nuclear system at an early time and quantum decoherence time of the nuclear system is of the order of $10^{-18}$ sec.

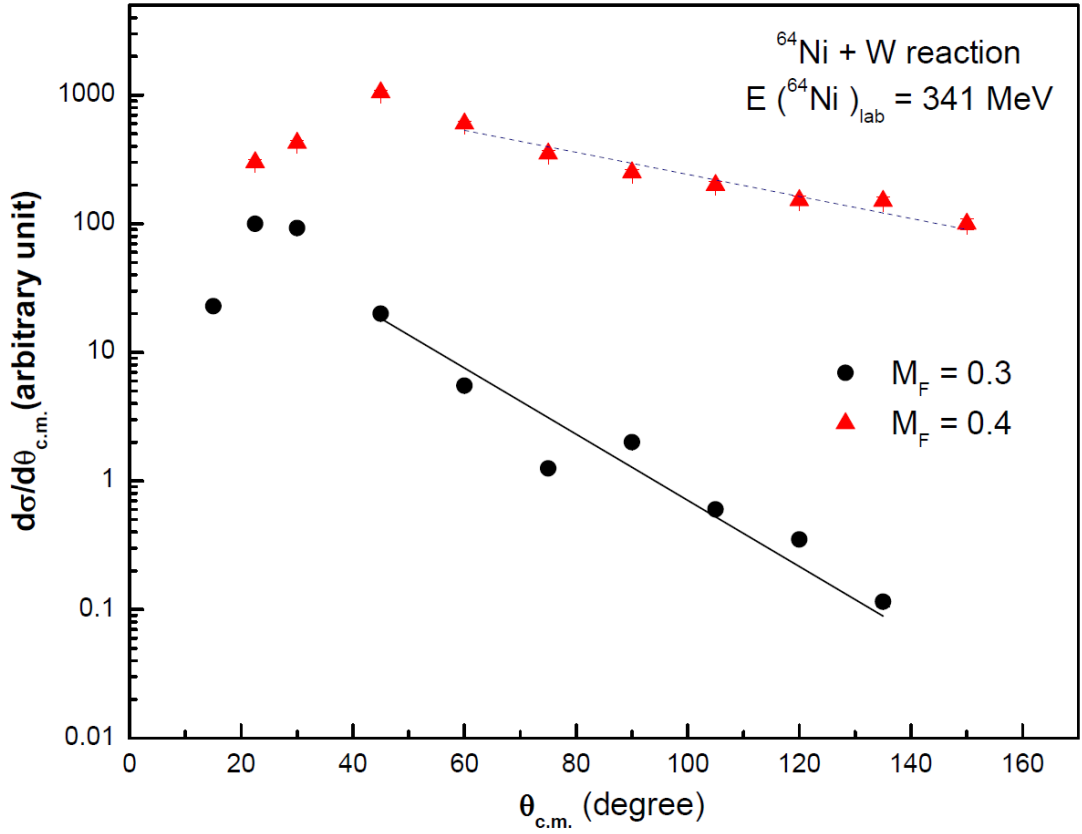

Fig. 1: (Color online) $\left(\frac{d\sigma}{d\theta}\right)_{c.m.}$ versus $\theta_{c.m.}$ plot for $^{64}$Ni+W reaction for M$_F$=0.3 and M$_F$=0.4.

17